\documentclass{hep99}
\usepackage{epsfig,cite}
\input axodraw.sty
\begin{document}
\input macros.sty

\title{%
\vspace{-3.1cm}
\begin{flushright}
       {\normalsize\tt OUTP-99-60-P}    \\[-0.2cm]
       {\normalsize\tt November 1999}   \\
\end{flushright}
       \vspace{0.7cm}
Lattice Gauge Theory$^\dag$}

\author{Hartmut Wittig}
\address{Theoretical Physics, Oxford University,
         1~Keble Road, Oxford OX1~3NP, UK}
\abstract{
  The status of lattice calculations in Quantum Field Theory is
  reviewed. A major part is devoted to recent progress in formulating
  exact chiral symmetry on the lattice.
  Another topic which has received a lot of attention is the influence
  of dynamical quark effects. Attempts to quantify these effects for
  the light hadron spectrum and flavour singlet amplitudes are
  discussed, as well as other expected qualitative features of
  simulations with dynamical quarks. The remaining parts of the review
  include recent results for the light quark masses using
  non-perturbative renormalisation and the spectrum of glueballs and
  heavy hybrids computed using anisotropic lattices.  
} 
\maketitle

\fntext{\dag}{Plenary talk presented at International Europhysics
  Conference on High-Energy Physics (EPS-HEP\,99), Tampere, Finland,
  15--21 July 1999.}

\section{Introduction}

Since Wilson's original formulation of lattice gauge theories in
1974\,\cite{Wilson74}, lattice methods have developed into a mature
area of research with a wide range of applications, including subjects
as diverse as QCD, Higgs models and Quantum Gravity. It is therefore
quite impossible to cover all recent activities in Lattice Gauge
Theory in a single review. Therefore, I shall concentrate on lattice
QCD applied to the calculation of hadron masses and the evaluation of
weak matrix elements. Another subject which I shall present in detail,
and which has received a lot of attention recently, is the lattice
formulation of chiral symmetry. Indeed, over the past two years it has
emerged that chiral gauge theories can be put on the lattice in a
consistent way, something which has been thought to be impossible for
a long time. These new developments have wide-ranging implications for
a large class of theories, which clearly merits a detailed discussion.

An overview of those areas which will {\it not\/} be discussed here
can be found in the proceedings of recent annual conferences on
Lattice Field Theory. In particular, I refer the reader to the plenary
talks on
\begin{itemize}
\item QCD at finite temperature and
      density\,\cite{edwin_lat97,alford_lat98,karsch_lat99}
\item The electroweak phase
      transition\,\cite{smit_lat97,mikko_lat98,fodor_lat99} 
\item Topology and confinement\,\cite{negele_lat98,mike_lat99}
\item Lattice gravity and random
      surfaces\,\cite{bowick_lat97,thor_lat98}
\end{itemize}
Furthermore there are reviews of lattice gauge theory given at recent
conferences on High Energy
Physics\,\cite{luscher_lp97,sharpe_ichep98}.

\subsection{General remarks}
\label{sec_general_intro}

Many of the concepts and techniques of Lattice Gauge Theories can be
easily introduced in the context of QCD. Here, the lattice formulation
provides a non-perturbative framework to compute relations between
Standard Model parameters and experimental quantities from first
principles.

The discretisation is achieved by introducing a euclidean space-time
lattice with spacing~$a$ and volume $L^3\cdot{T}$. The inverse lattice
spacing, $a^{-1}$ acts as an UV cutoff, which preserves the gauge
invariance of the theory. The quark and antiquark fields
$\psi(x),\,\psibar(x)$ are associated with the lattice sites~$x$,
whereas the gauge field is represented by the so-called link variable
$U_\mu(x)$, which connects neighbouring lattice sites, and is an
element of the gauge group SU(3). After choosing suitable
discretisations of the Yang-Mills action~$\Sg$ and the quark action
$\Sf$, the expectation value of an observable~$\Omega$ is defined as
\be
  \langle\Omega\rangle = \frac{1}{Z} \int\prod_{x,\mu}dU_\mu(x) \int\prod_x
  d\psibar(x)d\psi(x)\,\Omega\rme^{-\Sg-\Sf}
\ee
and the functional integral $Z$ is determined by requiring
$\langle\unt\rangle=1$. The discretisation procedure has hence given a
meaning to the functional integral measure, which becomes a simple
product measure. More importantly, after integrating out the quarks
this formulation allows for a {\em stochastic} evaluation of
$\langle\Omega\rangle$ using Monte Carlo techniques.

Ideally one would like to use lattice QCD as a phenomenological tool.
A typical application would be, for instance, the computation of the
strange quark mass $m_{\rm s}^\MSbar(2\,\gev)$, using the kaon mass,
$m_K$ and the pion decay constant as input parameters. However, as I
shall explain below, {\em realistic} simulations of lattice QCD are
difficult.

The first problem one has to address are lattice artefacts (cutoff
effects). Let $\Omega$ denote the quantity we wish to compute on the
lattice, e.g. a hadron mass. Then the expectation values on the
lattice and in the continuum differ by corrections of order $a^p$, viz
\be
  \langle\Omega\rangle^{\rm lat} = \langle\Omega\rangle^{\rm cont}
  +O(a^p), 
\label{eq_o_cont_lat}
\ee
where the power~$p$ in the correction term depends on the chosen
discretisation of the QCD action. Values of~$a$ which can currently be
simulated lie in the range $a\approx0.2-0.05\,\fm$. The size of the
correction term can in some cases be as large as $20$\,\%, depending
on the quantity and the chosen discretisation.
It is then clear that an extrapolation to the continuum, $a\to0$ is
required in order to obtain the desired result. This extrapolation can
be much better controlled if the chosen discretisation avoids small
values of~$p$.

Perhaps the biggest challenge as far as lattice QCD is concerned, is
the inclusion of dynamical quark effects. After integrating out the
quark fields the expression for the expectation value reads
\be
  \langle\Omega\rangle = \frac{1}{Z} \int\prod_{x,\mu}dU_\mu(x)
  \prod_f\det\left(D+m_f\right)\,\Omega\,\rme^{-\Sg},
\label{eq_quarkdet}
\ee
where $D$ is the lattice Dirac operator and $m_f$ is the mass of quark
flavour~$f$. The evaluation of the determinant in \eq{eq_quarkdet} in
numerical simulations is still very costly, even on today's massively
parallel computers. In many applications the determinant has therefore
been set to~1. This defines the so-called {\em quenched
approximation}, which corresponds to neglecting quark loops in the
evaluation of $\langle\Omega\rangle$. Although this represents a
rather drastic assumption about the influence of quark-induced quantum
effects, the quenched approximation works surprisingly well, as I
shall describe later.

An indirect consequence of using the quenched approximation is the
observed {\em scale ambiguity}. That is, the calibration of the
lattice spacing in physical units, $a^{-1}\,[\mev]$, is dependent on
the quantity $Q$ which is used to set the scale
\be
   a^{-1}\,[\mev] = \frac{Q\,[\mev]}{(aQ)},\quad Q=f_\pi, m_\rho,\ldots
\ee
This ambiguity arises because different quantities~$Q$ are affected by
quark loops in different ways.

There are also restrictions on the quark masses $\mq$ that can be
simulated. In general the following inequalities should be satisfied
\be
  a\ll\xi\ll L,
\ee
where~$L$ is the spatial extent of the lattice volume. The
quantity~$\xi$ denotes the correlation length of a typical hadronic
state, which serves as a measure of the quark mass. The inequality on
the right places restrictions on the light quark masses that can be
simulated: if those are too light one may suffer from finite-size
effects, since~$\xi$ becomes large. Typical spatial extensions of
$L\approx1.5-3\,\fm$ imply that the physical pion mass cannot be
reached. The left inequality restricts the masses of heavy quarks.
Since $a^{-1}\approx2-4\,\gev$, it is clear that relativistic
$b$-quarks cannot be simulated. One therefore relies on extrapolations
in $\mq$ to connect to the physical $u,\,d$ and~$b$ quarks. Obviously
it is of great importance to control such extrapolations.

Finally there is the issue of chiral symmetry breaking. A famous no-go
theorem by Nielsen and Ninomiya\,\cite{NiNi} implies that under fairly
mild assumptions exact chiral symmetry cannot be realised at non-zero
lattice spacing. Therefore, the chiral and continuum limits cannot be
separated. For many (but not all) applications of lattice QCD this may
not be a severe limitation, but the no-go theorem has so far precluded
all attempts to achieve a realistic lattice formulation of the
electroweak sector of the Standard Model. 

\subsection{Outline}

The remainder of this article is as follows: section~\ref{sec_dirac}
deals with the properties of lattice Dirac operators: it is described
how lattice artefacts in observables can be successfully reduced by
constructing ``improved'' lattice actions. The major part of
section~\ref{sec_dirac} is devoted to recent developments which have
ultimately led to the construction of lattice chiral gauge theories
with exact gauge invariance. Section~\ref{sec_dynquarks} discusses
recent simulations of QCD with dynamical quarks. Results which
quantify sea quark effects in the light hadron spectrum are presented,
and moreover the qualitative effects of sea quarks in flavour-singlet
amplitudes and regarding the breakdown of linear confinement (``string
breaking'') are discussed. In section~\ref{sec_quarkmass} recent
results for the light quark masses are presented. Progress in this
area has been achieved through the non-perturbative matching of
lattice results to the $\MSbar$ scheme. In section~\ref{sec_glueb} I
describe recent results for the spectrum of glueballs and heavy
hybrids.  Section~\ref{sec_other} contains a brief overview of results
for some weak hadronic matrix elements which could not be reviewed
extensively due to lack of time. Finally, a summary is presented in
section~\ref{sec_summary}.

\section{The lattice Dirac operator}
\label{sec_dirac}

This section deals with the general properties of lattice fermions.
After recalling the fermion doubling problem we shall discuss the
implementation of the Symanzik on-shell improvement programme, which
systematically reduces lattice artefacts in physical observables and
thus makes extrapolations to the continuum limit more reliable. The
main part of this section, however, reviews the recent progress made
in formulating chiral symmetry on the lattice.

\subsection{Fermion doubling revisited}

Suppose we want to describe massless {\em free} fermions on the
lattice. The lattice action can be written as
\be
  \Sf = a^4\sum_{x,y}\psibar(x)\,D(x-y)\,\psi(y),
\ee
where~$D$ denotes the lattice Dirac operator. In particular, one would
like to formulate the theory such that~$D$ satisfies the following
conditions 
\begin{itemize}
\itemsep 3pt
\item[(a)]      $D(x-y)$ is local 
\item[(b)]      $D(p)=i\gamma_\mu p_\mu +O(ap^2)$
\item[(c)]      $D(p)$ is invertible for $p\not=0$
\item[(d)]      $\gamma_5\,D+D\,\gamma_5=0$
\end{itemize}
Locality is required in order to ensure renormalisability and
universality of the continuum limit; it ensures that a consistent
field theory is obtained. Furthermore, condition~(c) ensures that no
additional poles occur at non-zero momentum. If this is not satisfied,
as is the case for the ``na\"{\i}ve'' discretisation of the Dirac
operator, additional poles corresponding to spurious fermion states
can appear: this is the famous fermion doubling problem. Finally,
condition~(d) implies that $\Sf$ is chirally invariant.

The main conclusion of the Nielsen-Ninomiya no-go theorem is that
conditions~(a)--(d) cannot be satisfied simultaneously. Since one is
not willing to give up locality and condition~(b), this implies that
one is usually confronted with the choice of tolerating either doubler
states or explicit chiral symmetry breaking. This is manifest in the
two most widely used lattice fermion formulations: staggered
(``Kogut-Susskind'') fermions leave a chiral U(1) subgroup invariant,
but only partially reduce the number of doubler species~\cite{KS75}.
Wilson fermions, on the other hand, remove the doublers entirely at
the expense of breaking chiral symmetry explicitly. This is easily
seen from the expression for the free Wilson-Dirac operator
\be
  D_W^{(0)} = \textstyle\frac{1}{2}
             \gamma_\mu\left(\nabla_\mu+\nabla_\mu^*\right) 
             -\textstyle\frac{1}{2}a\nabla_\mu^*\nabla_\mu.
\ee
Using the definitions for the forward and backward lattice
derivatives, $\nabla_\mu$ and $\nabla_\mu^*$, one easily proves
conditions~(a)--(c), while it is obvious that~(d) is not satisfied.
Nevertheless, the (interacting) Wilson-Dirac operator~$D_W$ is widely
used in simulations of lattice QCD. This is because chiral symmetry
breaking in a vector-like theory is often merely an inconvenience, but
no fundamental obstacle.

\subsection{Improved discretisations of the Wilson action}

Another consequence of using the Wilson-Dirac operator is the presence
of large cutoff effects in physical observables. One can show that the
leading lattice artefacts are of order~$a$ (i.e. $p=1$
in~\eq{eq_o_cont_lat}), while e.g. for staggered fermions the lattice
corrections start at order~$a^2$.

It is important to realise the definition of a lattice action is not
unique: one can add any number of operators which formally vanish in
the continuum limit, provided that they comply with the correct
symmetry and locality requirements. Given this relative freedom in
defining lattice actions, an obvious question is whether one can find
{\it improved\/} discretisations for lattice fermions, for which the
cutoff effects are reduced. Two approaches have been the subject of
much recent activity. One is based on the Symanzik improvement
programme~\cite{SymanzikI,SymanzikII}, in which lattice artefacts are
removed order by order in the lattice spacing. In the second approach
one seeks to construct a ``perfect'' lattice action (which is
essentially free of lattice artefacts), using renormalisation group
techniques~\cite{HasNie_perf}.

Sheikholeslami and Wohlert have shown that in QCD with Wilson fermions
the Symanzik improvement programme can be implemented to lowest order
in~$a$ by adding one dimension-5 counterterm to~$D_W$. The resulting
operator (with bare mass $m_0$) reads
\be
   D_{SW} = D_W +m_0
   +\textstyle\frac{ia}{4}\,\csw\sigma_{\mu\nu}F_{\mu\nu},
\label{eq_DSW}
\ee
where $F_{\mu\nu}$ is a lattice transcription of the field tensor. In
order to remove all lattice artefacts of order~$a$ in hadron masses,
the improvement coefficient~$\csw$ has to be fixed by imposing a
suitable improvement condition. Such a condition is provided by
requiring that the restoration of the axial Ward identity holds up to
terms of order~$a^2$~\cite{alphaI}. It has been applied to
determine~$\csw$ non-perturbatively for a large range of couplings for
both quenched QCD~\cite{alphaIII,SCRI_imp} and also for $\nf=2$
flavours of dynamical quarks~\cite{JansenSommer_98}.

In order to compute matrix elements of local composite operators such
as vector and axial vector currents, one has to consider their
correctly normalised and improved versions. For instance, the
renormalised axial current $(A_{\rm R})_\mu$ in the $O(a)$ improved
theory reads
\be
  (A_{\rm R})_\mu =
  \za(1+\ba a\mq)\left\{A_\mu+\ca a\partial_\mu P\right\}
\ee
where $\ba$ and $\ca$ are improvement coefficients, $P$ is the
pseudoscalar density, and $\za$ is the renormalisation constant for
the axial current. In refs.~\cite{alphaIII,alphaIV} $\ca$ and $\za$
have been determined non-perturbatively, and a general strategy to
determine the improvement coefficients and renormalisation constants
of all quark bilinears in the $O(a)$ improved theory has been
presented in~\cite{gupta_sharpe_etal}.  Detailed studies of the matrix
elements of such operators with much better controlled lattice
artefacts should therefore soon be feasible.

\begin{figure}[tb]
\vspace{-1.0cm}
\ewxy{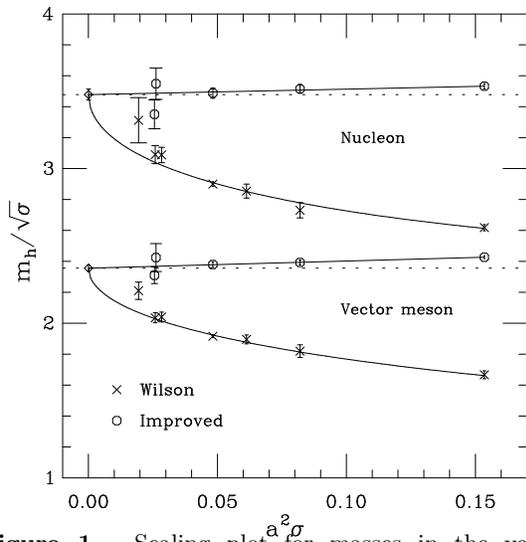}{90mm}
\vspace{-1.2cm}
\caption{Scaling plot for masses in the vector meson and nucleon
  channels for improved (circles) and unimproved (crosses) Wilson
  fermions~\protect\cite{SCRI_imp}. The linearity of the data computed
  with the improved action demonstrates the expected scaling
  behaviour.
}
\label{fig_scaling}
\end{figure}

The remaining question is whether non-perturbative $O(a)$ improvement
can be {\em verified\/}, in the sense that physical observables
computed in the $O(a)$ improved theory can be shown to approach their
continuum limit with a rate proportional to~$a^2$~\cite{how_lat97}.
Detailed scaling studies performed in the quenched approximation have
shown that this is indeed the case for meson and baryon
masses~\cite{SCRI_imp}, as well as for matrix elements of vector and
axial vector currents~\cite{jochen99}.  An example is shown in
Fig.~\ref{fig_scaling}, taken from ref.~\cite{SCRI_imp}.

In ref.~\cite{AlfKlaLep} the Symanzik improvement programme was
extended to higher orders. The many relevant improvement coefficients
have been determined in (mean-field improved~\cite{lepenzie93})
perturbation theory. $O(a)$ improvement has also been studied for the
anisotropic Wilson action~\cite{klassen_lat98}, for which the temporal
and spatial lattice spacings are chosen differently. This offers an
advantage in the computation of heavy states, and we will return to
this point in section~\ref{sec_glueb}.

Another method to construct a lattice action with improved scaling
behaviour was presented in ref.~\cite{DeGHasKov99}. The idea is to use
the operator~$D_{SW}$ in \eq{eq_DSW} on smoothed gauge
configurations. Such configurations are obtained through a blocking
procedure applied to spatial link variables as described
in~\cite{albanese87}. The resulting action is called the ``fat-link''
clover action. In a series of papers~\cite{DeGrand99,SteDeDeHa99} it
has been shown that fat-link clover actions exhibit good chiral
properties: the additive renormalisation of the quark mass encountered
for Wilson fermions is small (yet non-vanishing), and the
renormalisation factors $\za$ and~$\zv$ of axial and vector currents
are close to unity. Furthermore, scaling tests of various hadron
masses have been carried out to test whether cutoff effects of
order~$a$ have been eliminated.

One concludes that improved discretisations of the Wilson actions are
successful and play an ever more important r\^ole in calculations of
the hadron spectrum and matrix elements. Their main advantage is that
more accurate results in the continuum limit are obtained. It is
important to keep in mind, though, that improved discretisations
obtained through the Symanzik improvement programme and also the
fat-link clover actions do not alleviate the problem of explicit
chiral symmetry breaking.

\subsection{Exact chiral symmetry on the lattice}

A lot of progress has been made recently in the formulation of chiral
fermions on the lattice. In fact, it has been shown how lattice chiral
gauge theories can be formulated in a way which preserves locality and
gauge invariance. Many new developments in this field have followed
the rediscovery of the Ginsparg-Wilson relation~\cite{GW82}
\be
   \gamma_5\,D+D\,\gamma_5 = aD{\gamma_5}D.
\label{eq_GW}
\ee
The significance of this relation had not been realised for a long
time, since no non-trivial expression for a Dirac operator~$D$
satisfying~\eq{eq_GW} was known. It is remarkable that two
constructions of such a solution have been developed independently at
around the same time.

The first goes back to Kaplan's proposal to realise chiral fermions in
the domain wall fermion approach~\cite{kaplan92}. This was
subsequently re-cast by Narayanan and Neuberger, who developed the
``overlap'' representation of the chiral
determinant~\cite{NarNeu_ovlp1,NarNeu_ovlp2}. Furthermore, Shamir and
Furman~\cite{ShaFur94} used a variant of Kaplan's work to derive the
domain wall fermion formulation for vector-like theories like
QCD~\cite{ShaFur94}. The second solution to~\eq{eq_GW} was constructed
in the perfect action approach mentioned earlier.

Here I will briefly discuss some of the concepts and applications.
More details can be found in recent reviews on Domain Wall
fermions~\cite{blum_lat98} and the construction of lattice chiral
gauge theories~\cite{neu_lat99,martin_lat99}.

I shall first discuss the Domain Wall (DW) fermion formulation as
described in ref.~\cite{ShaFur94}. The basic idea, proposed
in~\cite{kaplan92} is to introduce an extra (fifth) dimension and to
consider fermions coupled to a mass defect in the extra dimension.  To
make this more explicit, let~$x,y$ denote 4-dim. coordinates and
$s,s^\prime$ the coordinates in the $\rm 5^{th}$ dimension, which has
finite length $N_s$. The gauge fields are trivial in the $\rm 5^{th}$
direction, and the Dirac operator then has the form
\be
  D_{ss^\prime}^{\rm DWF}(x,y) = D^\parallel(x,y)\delta_{ss^\prime} +
    \delta(x-y)D_{ss^\prime}^\perp
\ee
where $D^\parallel(x,y)$ is the usual Wilson-Dirac operator with a
{\it negative\/} mass term, $-M$. The operator $D_{ss^\prime}^\perp$
couples fermions in the extra dimension and contains the Dirac
mass~$m$. It can now be shown that for $m=0$ and in the limit
$N_s\to\infty$ there are no fermion doublers and, more importantly,
chiral modes of opposite chirality are trapped in the 4-dim. domain
walls at $s=1,\,N_s$. The physical, 4-dim. fields defined by
\bea
  q(x) = P_{\rm R}\psi_1(x) +P_{\rm L}\psi_{N_s}(x) \nonumber\\
  \overline{q}(x) = \psibar_{N_s}(x)P_{\rm R} +\psibar_1(x)P_{\rm L}
\eea
satisfy exact, continuum-like axial Ward identities at non-zero values
of~$a$ for $N_s\to\infty$. This result, derived in~\cite{ShaFur94},
formally establishes the correct chiral properties of the lattice
regularised theory.

In a real simulation one has to work at finite~$N_s$ so that the
decoupling of chiral modes is not exact. It is possible to show,
however, that the terms which break the chiral symmetry are
exponentially suppressed. This point can be illustrated by discussing
the relation between the pion mass and the quark mass in Domain Wall
QCD. At finite~$a$ one has
\be
  (am_\pi)^2 = C\left(am+a\rme^{-\gamma{N_s}}\right),
\label{eq_pion_dwf}
\ee
where~$C$ is a constant, $m$ is the Dirac mass which appears in
$D^\perp$, and $\gamma>0$. This expression shows that in the limit
$N_s\to\infty$ the bare quark mass is only multiplicatively
renormalised (i.e. there is no additive quark mass renormalisation as
for ordinary Wilson fermions). It also shows that for finite~$a$ there
is a residual pion mass in the chiral limit proportional to
$a\exp(-\gamma{N_s})$.

Hence, the Domain Wall formulation of QCD offers a method to realise
almost exact chiral invariance at non-zero lattice spacing at the
expense of simulating a 5-dim. theory. An important feature is that
lattice artefacts of order~$a$ are exponentially suppressed at
finite~$N_s$. In other words, the theory is $O(a)$ improved for
$N_s\to\infty$, with exponentially small $O(a)$ corrections expected
at finite~$N_s$. As in the case of non-perturbative $O(a)$ improvement
the expected improved scaling behaviour in Domain Wall QCD must be
verified.

\begin{figure}[tb]
\vspace{-2.5cm}
\ewxy{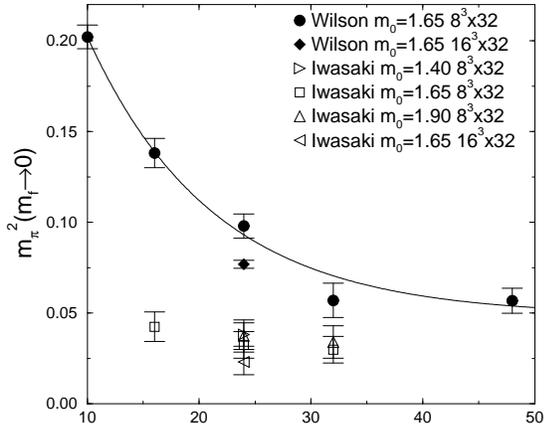}{90mm}
\vspace{-1.0cm}
\caption{The pion mass in the chiral limit plotted versus the size of
the 5th dimension from ref.~\protect\cite{CU_Wu_lat99}. Solid
circles are the data obtained using the Wilson plaquette action for
the gauge field, whereas open symbols denote data obtained using an
improved gauge action (Iwasaki). The solid line is a fit assuming
$(am_\pi)^2=A+B\exp(-\gamma{N_s})$, which yields a non-vanishing
intercept of $A=0.048(9)$ in the limit $N_s\to\infty$ (notation
translates as: $L_s\to N_s, m_0\to{M}, m_f\to{m}$).}
\label{fig_mres_Nsdep}
\end{figure}

The question that arises is how small a value of~$N_s$ one can get
away with in order to realise chiral symmetry whilst keeping the
computational overhead of simulating an extra dimension at a
minimum. In refs.~\cite{CU_Wu_lat99,CU_Flem_lat99} systematic studies
of the $N_s$-dependence of $(am_\pi)^2$ in the chiral limit have been
presented. An example of such an analysis is shown in
Fig.~\ref{fig_mres_Nsdep}~\cite{CU_Wu_lat99}. The main conclusion is
that the suppression of the residual pion mass expected according to
\eq{eq_pion_dwf} is very slow. At present it is not even clear that it
vanishes at all in the limit $N_s\to\infty$. Further studies,
clarifying the r\^ole of conventional, 3-dim. finite-size effects and
the effects of quenching, are required to settle this issue.

There are already extensive numerical studies of physical observables
using DW\,QCD. Results include studies of QCD
thermodynamics~\cite{Colum_th}, weak interaction matrix elements such
as those relevant for kaon physics, i.e. the $B$-parameter $B_K$ and
$\epsilon^\prime/\epsilon$~\cite{BluSo_BK,BluSo_lat99,BRC_eps} and
determinations of the light quark masses~\cite{BSW_quark,Wing_lat99}.

Turning now to the question of constructing chiral gauge theories on
the lattice with exact gauge invariance, let us recall that a central
ingredient in the construction of such theories is the Ginsparg-Wilson
(GW) relation~\eq{eq_GW}, which replaces the condition
$\{\gamma_5,D\}=0$. One key observation, made in~\cite{HasLalNie98},
is that any lattice Dirac operator satisfying the GW relation has
chiral zero modes and satisfies an exact index theorem. It was
subsequently realised~\cite{martin_ch1} that the GW relation implies
an exact symmetry of the associated action, with infinitesimal
variations proportional to
\bea
  & &\delta\psi=\gamma_5(1-\textstyle\frac{1}{2}aD)\psi \nonumber\\
  & &\delta\psibar=\psibar(1-\textstyle\frac{1}{2}aD)\gamma_5.
\eea
Moreover, this symmetry reproduces the correct chiral anomaly in the
flavour singlet case. That is, all the hallmarks of the correct chiral
behaviour are present in the lattice theory: chiral zero modes, an
exact index theorem and the chiral anomaly derived from the Ward
identities associated with the exact symmetry.

At this stage left- and right-handed fermions are easily
introduced. Furthermore, starting from a solution to the GW relation,
it is possible to construct abelian chiral gauge theories on the
lattice, which comply with all basic requirements, including exact
gauge invariance~\cite{martin_ch3}. The construction extends to the
non-abelian case~\cite{martin_ch4}, and although a few properties have
yet to be established with the same rigour as for abelian theories,
there is little doubt that the construction is valid.

An operator which satisfies the GW relation can be constructed in the
framework of the overlap formalism. After Kaplan's proposal and
independent related work by Frolov and Slavnov~\cite{FroSla93},
Narayanan and Neuberger obtained an expression for the chiral
determinant, which is usually termed the
``overlap''~\cite{NarNeu_ovlp1,NarNeu_ovlp2}. Among many other things
the overlap was shown to reproduce the correct chiral properties in
vector-like theories like QCD~\cite{NarNeu_ovlp2}. Furthermore, it can
be brought into a more manageable expression in terms of a relatively
simple lattice Dirac operator~$D_{\rm N}$. It is defined
by~\cite{Neu_op98_1}
\bea
  & & D_{\rm N} = {\textstyle\frac{1}{2}}\left(
  1-X({X^\dagger}X)^{-1/2}\right) \nonumber\\
  & & X=1-D_W,
\eea
where $D_W$ is the massless Wilson-Dirac operator. $D_{\rm N}$ can
easily be shown to satisfy the GW relation~\cite{Neu_op98_2}. Moreover,
it was demonstrated that the construction of $D_{\rm N}$ from the
overlap could be reversed, using only the GW relation and
factorisation~\cite{Nar98}.

The obvious question at this point is whether the correct chiral
properties of operators satisfying the GW relation can be verified in
lattice simulations. A possible strategy is then to check whether the
expected global anomalies are exhibited. One example is Witten's
observation~\cite{Witten82} that SU(2) gauge theory coupled to a
single left-handed fermion is mathematically inconsistent. Restricting
the discussion to the continuum theory for the moment, this is
easily seen by examining the functional integral for this theory after
the Weyl fermions have been integrated out, viz
\be
  {\cal Z}=\int [dA]\,\left(\det D[A] \right)^{1/2}\,\rme^{-\Sg[A]}.
\label{eq_ambig}
\ee
Here, $A$ is the SU(2) gauge field, and $D$ denotes the Weyl operator.
Obviously a sign ambiguity arises due to the presence of the square
root in~\eq{eq_ambig}. Indeed, for SU(2) there exist non-trivial gauge
transformations $g(x)$ which cannot be deformed continuously to the
identity, and for which
\be
  \left(\det D[A] \right)^{1/2} = -\left(\det D[A^g] \right)^{1/2},
\ee
where $A^g$ denotes the transformed gauge field. This result implies
that expectation values defined through ${\cal Z}$ are indeterminate,
i.e. $\langle\Omega\rangle =$``0/0''.

Although $A_\mu$ and $A_\mu^g$ are not connected via some smooth gauge
transformation, they are none the less connected in the space of gauge
fields. Thus, one can define a curve which smoothly interpolates
between $A_\mu$ and $A_\mu^g$, given by
\be
  A_\mu(t) = (1-t)\,A_\mu + t A_\mu^g,\quad t\in[0,1].
\label{eq_Amu_curve}
\ee
By invoking the Atiyah-Singer index theorem, Witten showed that the
number of eigenvalues of the Dirac operator which cross zero is
odd. This observation about the behaviour of the spectral flow along
the curve $A_\mu(t)$ then gives rise to the sign ambiguity and hence
the mathematical inconsistency. It is now interesting to investigate
whether the expected spectral flow can be reproduced on the lattice.

\begin{figure}[tb]
\ewxy{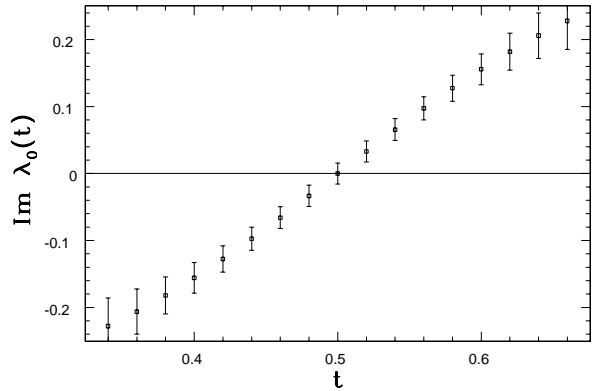}{90mm}
\vspace{-3.8cm}
\caption{Level crossing for the imaginary part of the lowest eigenvalue
of the overlap operator}
\label{fig_imlambda}
\end{figure}

Initially this has been investigated by Neuberger~\cite{Neu_cross}.
Then, in ref.~\cite{BaerCam_lat99}, B\"ar and Campos reported on the
computation of the eigenvalues of the overlap operator~$D_{\rm N}$ in
a lattice simulation along the path which smoothly connects a constant
gauge configuration to its gauge transform as in~\eq{eq_Amu_curve}.
Focussing on the six lowest eigenvalues they found that only one, i.e.
the smallest, $\lambda_0$, becomes zero at $t=0.5$. The hermiticity
properties of~$D_{\rm N}$ imply that one expects the level crossing to
occur for the imaginary part of this eigenvalue.
Figure~\ref{fig_imlambda} demonstrates that the crossing is indeed
observed for $\Im\lambda_0$. This result represents a numerical proof
of Witten's original argument and is only possible if the lattice
Dirac operator has the correct chiral properties. The above example
illustrates the enormous progress that has been achieved in the
formulation of chiral symmetry.

Other recent work in this area includes detailed studies of the axial
anomaly~\cite{KikuYam_98,Chiu,Fujikawa98,Suzuki98,Adams98,ReiRoth99},
investigations of spontaneous chiral symmetry
breaking~\cite{EdHelNar_cc,DamEdHelNar,HerJanLel99}, the locality
properties of $D_{\rm N}$~\cite{HerJanLue} and the development of
efficient implementations of Ginsparg-Wilson
fermions~\cite{EdHelNar_alg,KenHorSin,Borici,McIrMi,DeGrand_GW}.

In view of these results it should be clear now that a consistent
formulation of the Standard Model exists beyond perturbation
theory. There are also important consequences for future lattice
studies of supersymmetric models: in many ways the intrinsically
supersymmetric features in such models bear resemblance to the r\^ole
of chiral symmetry in QCD.

\section{Simulations with dynamical quarks}
\label{sec_dynquarks}

Now we turn to discuss the other great challenge in present
simulations of QCD, namely the inclusion of dynamical quark effects.
As mentioned in section~\ref{sec_general_intro} the quenched
approximation is still widely used for many phenomenologically
interesting quantities, and in order to enhance the predictive power
of lattice simulations it is of great importance to assess the
influence of dynamical quarks.

Before such an assessment can be made with the required accuracy, it
is necessary to perform precise calculations of experimentally known
quantities in the quenched theory. Such benchmarks serve not only to
detect significant effects due to dynamical quarks, but also
illustrate how closely the quenched approximation resembles the real
world.

\subsection{Quenched light hadron spectrum}

\begin{figure}[tb]
\vspace{1.cm}
\ewxy{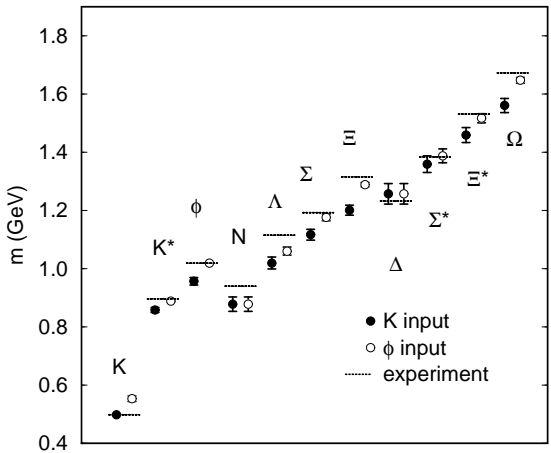}{90mm}
\vspace{-4.0cm}
\caption{The quenched hadron spectrum from
ref.~\protect\cite{CP-PACS_quen} compared to experiment (dashed lines).}
\label{fig_CPPACS_quen}
\end{figure}

The CP-PACS Collaboration recently presented a precision calculation
of the light hadron spectrum in quenched QCD using the (unimproved)
Wilson action~\cite{CP-PACS_quen}, which superseded earlier studies by
GF11~\cite{GF11_quen}. The findings of CP-PACS are summarised in the
plot shown in Fig.~\ref{fig_CPPACS_quen}. Although the qualitative
features of the spectrum are well reproduced by the quenched lattice
data, one finds significant deviations from the experimentally
observed spectrum. For instance, the ratio of the nucleon and $\rho$
masses is calculated as
\be
  m_N/m_\rho = 1.143\pm0.033,
\ee
which is 6.7\,\% (2.5\,$\sigma$) below the experimental value
of~1.218. Similarly, vector-pseudoscalar mass splittings such as
$m_{K^*}-m_K$ are too small by $10-16$\,\% ($4-6\,\sigma$), depending
on whether $m_K$ or $m_\phi$ is used as input to fix the strange quark
mass. This result implies that, for the first time, a {\it
  significant\/} deviation between the quenched QCD spectrum and
nature is detected. The main conclusion of ref.~\cite{CP-PACS_quen} is
that quenched QCD describes the light hadron spectrum at the level
of~10\,\%. However, it also shows that the quenched approximation
works surprisingly well, since the discrepancy is fairly mild. This
has important consequences for other, phenomenologically more
interesting quantities, for which one may be stuck with the quenched
approximation for some time to come.

Although the CP-PACS results represent a real benchmark in terms of
statistics, parameter values and lattice volumes, further
corroboration of these findings is still required. It should be added
that the results depend crucially on the modelling of the quark mass
dependence of observables. Usually the results of Chiral Perturbation
Theory can be used to guide the extrapolations in the quark masses.
However, in the quenched approximation one expects deviations from the
predictions of Chiral Perturbation Theory due to the appearance of
``quenched chiral logarithms''~\cite{BerGol92,Sharpe92}, which makes
the extrapolation of hadron masses close to the chiral limit hard to
control. Furthermore, the CP-PACS results were obtained using
unimproved Wilson fermions which have large lattice artefacts. It is
therefore desirable to check whether the extrapolations to both the
chiral and continuum limits are controlled.

Recent calculations employing different discretisations have largely
confirmed the findings of~\cite{CP-PACS_quen}: the MILC
Collaboration~\cite{MILC_quen} has used staggered fermions and finds a
value for the nucleon to rho mass ratio in the continuum limit of
\be
  m_N/m_\rho = 1.254\pm0.018\,{\rm(stat)}\pm0.028\,{\rm (syst)},
\ee
This is in broad agreement with experiment, but the difference to the
CP-PACS result amounts to only two~$\sigma$. A recent calculations by
UKQCD using $O(a)$ improved Wilson fermions~\cite{UKQCD_quen}
essentially confirms the conclusion of CP-PACS, namely that the
quenched light hadron spectrum agrees with experiment at the level
of~10\,\%.

\subsection{Light hadron spectrum for $\nf=2$}

An obvious question is whether sea quark effects can account for the
observed deviation of the quenched light hadron spectrum from
experiment. Before discussing some results it is useful to explain a
few technical issues which are relevant for simulations with dynamical
quarks.

First it is evident that improved lattice actions have an even more
important r\^ole to play. Since simulations with dynamical quarks are
very expensive, one cannot afford to control the extrapolation to the
continuum limit by simulating very small lattice spacings, whilst
keeping sufficiently large spatial volumes in physical units. In order
to be able to separate sea quark effects from lattice artefacts, it is
vital to have control over the latter. Details of recent simulations
with $\nf=2$ flavours of dynamical quarks are listed in
Table~\ref{tab_simpar_nf2}. The quark actions used in these
calculations are based on the $O(a)$ improved Wilson-Dirac
operator~$D_{SW}$ (using either the non-perturbative determination for
$\csw$ or its estimate in mean-field improved perturbation theory),
the unimproved operator~$D_W$, domain wall quarks or staggered
fermions.  In addition to the widely used plaquette action for the
gauge fields (labelled ``Plaq.'' in Table~\ref{tab_simpar_nf2}), two
collaborations also use improved gauge actions. Such actions were
introduced by L\"uscher and Weisz~\cite{LueWei85} and
Iwasaki~\cite{Iwasaki_imp}: they have been found to lead to smaller
lattice artefacts in the static quark potential, although the bulk of
the discretisation effects in hadronic quantities can only be reduced
by employing improved fermion actions~\cite{CP-PACS_comp}.

\begin{table}
\begin{center}
\caption{Recent simulations with $\nf=2$ flavours of dynamical quarks,
together with the sustained computer power (in GFlops). The other two
columns denote the choice of discretisation for the gauge action $\Sg$
and the Dirac operator used in $\Sf$.}
\label{tab_simpar_nf2}
\begin{tabular}{llll}
\hline\hline\\[-0.5ex]
Collab. & GFlops & $\Sg$ & $\Sf$ \\[0.5ex]
\hline\\[-0.5ex]
RBC     & $\sim250$     &       Plaq.   &       $D^{\rm DWF}$ \\
CP-PACS & $\sim300$     &       Iwasaki &       $D_{SW},\,\rm tad.$ \\
UKQCD   & $\sim{\0}30$  &       Plaq.   &       $D_{SW},\,\rm n.p.$ \\
SESAM/T$\chi$L & $\sim{\0}14$ & Plaq.   &       $D_{W}$ \\
MILC    & $\gtaeq\0\0{7}$ &       LW    &       $D_{\rm stag}$ \\[0.5ex]
\hline\hline
\end{tabular}
\end{center}
\end{table}

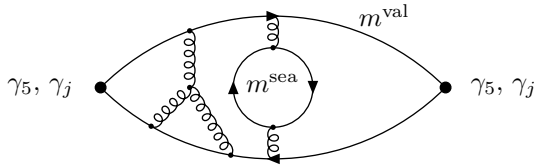
\begin{figure}[h] 
\begin{center} 
\vspace{-1.5cm}
\begin{picture}(150,150)(0,0) 
\thicklines 
\GCirc(10.0,75.00){2}{0}
\GCirc(140.,75.00){2}{0}
\put(70,72){\makebox(10,10){$\msea$}}
\put(117,97){\makebox(10,10)[tr]{$\mval$}}
\put(-10,70){\makebox(10,10)[r]{$\gamma_5,\,\gamma_j$}}
\put(150,70){\makebox(10,10)[l]{$\gamma_5,\,\gamma_j$}}
\ArrowArcn(75,10.00)(91.92,135,45) 
\ArrowArcn(75,140.0)(91.92,315,225)
\ArrowArcn(75,75)(15,270,90) 
\ArrowArcn(75,75)(15,90,270)
\Gluon(75,101.92)(75,90){2}{2} 
\Vertex(75,90){1}
\Gluon(75,48.08)(75,60){2}{2} 
\Vertex(75,60){1} 
\Vertex(43.56,96.38){1}
\Gluon(43.56,96.38)(43.56,75.00){2}{4}
\Vertex(43.56,75.00){1}
\Gluon(43.56,75.00)(29.04,60.39){2}{4}
\Vertex(29.04,60.39){1}
\Gluon(43.56,75.00)(59.04,49.48){2}{6}
\Vertex(59.04,49.48){1}
\end{picture}
\vspace{-1.5cm} 
\end{center}
\caption{Quark and gluon contributions to the two-point function for
  a pseudoscalar (vector) meson in partially quenched QCD with
  $\msea\ne\mval$.}
\label{fig_2pt_nf2}
\end{figure}

The two flavours of (degenerate) dynamical quarks with mass~$\msea$
are usually identified with the physical~$u$ and~$d$ quarks. Unlike in
the real world it is possible in lattice simulations to compute
observables relating to hadrons whose valence quarks have a different
mass than the sea quarks, i.e. $\mval\ne\msea$. This is illustrated in
the diagram in Fig~\ref{fig_2pt_nf2} for the case of a mesonic
two-point function. It implies that one has more freedom to explore
the dependence of physical observables on $\mval$ and~$\msea$
separately, and to compare the results to the predictions to Chiral
Perturbation Theory~\cite{GaLeu85} and its modified version for
unequal sea and valence quark
masses~\cite{Sharpe_PQ,GolLeu_PQ,BerGol_stag_PQ}.

It is clear that the simulations listed in Table~\ref{tab_simpar_nf2}
do not have the correct value of~$\nf$ for kaon physics. In the
absence of any sufficiently tested simulation algorithm capable of
treating $\nf=3$ flavours of dynamical fermions one is forced to
introduce the strange quark as a valence quark by making the
identifications
\be
  \ms=\mval,\quad m_{\rm u,d}=\msea,\quad \mval>\msea.
\ee
This defines the so-called {\it partially quenched} approximation.

One should also bear in mind that the values of~$\msea$ which are
accessible in current simulations (especially for Wilson quarks) are
still relatively large. This is easily seen by computing the ratio of
the pseudoscalar to vector meson mass~$\mps/\mv$ (for $\msea=\mval$)
and comparing it to the physical ratio $m_\pi/m_\rho=0.169$. For some
of the simulations in Table~\ref{tab_simpar_nf2} one
finds~\cite{SESAM_nf2_98,UKQCD_c176,joyce_lat99,CP-PACS_comp,kaneko_lat99}
\be
\frac{\mps}{\mv} = \left\{\begin{array}{lll}
0.69-0.83 && {\rm SESAM} \\
0.58-0.86 && {\rm UKQCD} \\
0.60-0.80 && {\rm \hbox{\rm CP-PACS}} \\
                          \end{array}\right.
\ee
In order to make contact with the physical situation one has to study
the dependence of observables on $\msea$ and extrapolate in~$\msea$ to
the physical value of $m_\pi/m_\rho$. Since the vector meson can decay
into two pseudoscalars below $\mps/\mv\approx0.5$ such a naive
extrapolation may, however, be misleading.

\begin{figure}[tb]
\vspace{1.cm}
\ewxy{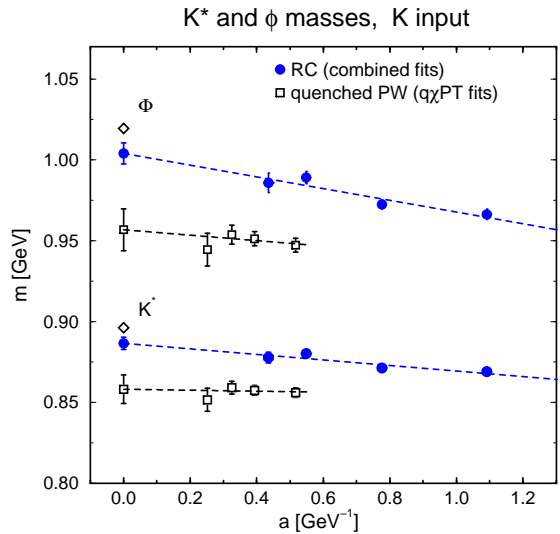}{90mm}
\vspace{-3.3cm}
\caption{Scaling behaviour of vector mesons in quenched and partially
quenched QCD~\protect\cite{kaneko_lat99}.}
\label{fig_vector_PQ}
\end{figure}

Several collaborations have investigated sea quark effects in the
light hadron spectrum. CP-PACS~\cite{kaneko_lat99} studied the
continuum limit of hadron masses computed for $\nf=2$ and compared it
with the results of the quenched light hadron spectrum discussed
before. Figure~\ref{fig_vector_PQ} shows the results for the masses of
the $K^*$ and~$\phi$ mesons for~$\nf=0,\,2$. This demonstrates clearly
that the discrepancy with the experimentally observed spectrum is
considerably reduced when sea quarks are ``switched on''. The figure
also illustrates that a reliable quantification of sea quark effects
can only be made after the extrapolation to the continuum limit: at
non-zero values of~$a$ the deviation between experiment and partially
quenched QCD is enhanced, and thus one would come to the wrong
conclusion about the size of dynamical quark effects.

Results for the vector-pseudoscalar mass splitting reported by
UKQCD~\cite{UKQCD_c176} also show that lattice data for this quantity
approach the experimental value as the sea quark mass is decreased.

These findings are quite encouraging: they demonstrate that sea quarks
have the expected effects on hadronic quantities. Of course, the
remaining differences between partially quenched QCD and experiment
seen in Fig.~\ref{fig_vector_PQ} have to be explained. It is
reasonable to assume that the incorrect value of~$\nf$ for kaon
physics will have some influence. Further studies are required in
order to decide whether there is yet sufficient control over the
extrapolations in $\msea$, $\mval$, as well as the extrapolations to
the continuum limit.

\subsection{Sea quark effects in other observables}

There are a number of observables which are particularly sensitive to
sea quark effects. One example are flavour-singlet amplitudes which
are relevant for quantities such as the $\eta^\prime$ mass. Unlike
flavour non-singlet matrix elements these amplitudes receive
contributions from disconnected diagrams and are thus sensitive to
vacuum polarisation effects due to dynamical quarks and the value
of~$\nf$. 

In a number of recent papers calculations of disconnected diagrams
contributing to the $\eta^\prime$ mass~\cite{KilVen,ruedi_lat99}, the
$\pi$-nucleon $\sigma$-term~\cite{SESAM_piN} and the flavour-singlet
axial coupling of the proton~\cite{SESAM_GA} have been presented.
Although some of the expected qualitative features of sea quark
contributions to these observables have been seen~\cite{ruedi_lat99},
the main obstacles for more quantitative analyses are the high level
of statistical noise encountered in the evaluation of disconnected
diagrams and also the fact that $\nf=3$ flavours cannot be simulated.
It is estimated that the calculation of the disconnected contribution
to the mass of flavour-singlet mesons is an order of magnitude more
expensive than conventional correlators, so that more efficient
numerical methods have to be developed and applied~\cite{cmi_disc}.
Therefore, there are no firm quantitative results for vacuum
polarisation in flavour-singlet amplitudes at present. The current
status of this field has been reviewed extensively
in~\cite{gusken_disc}.

A phenomenon which is expected to occur in the presence of dynamical
quarks is the breakdown of linear confinement, also called ``string
breaking''. This should manifest itself in a flattening-off of the
static quark potential at a characteristic distance~$r_{\rm b}$ at
which the mass of two heavy-light mesons is energetically favoured
over the energy of the flux tube which is responsible for the linearly
rising potential between the static quarks. In QCD string breaking can
be linked to decay rates of processes like
$\Upsilon(4{\rm{S}})\to{B}\overline{B}$~\cite{DruHor_98} and is
therefore of direct phenomenological relevance. Despite many recent
efforts~\cite{UKQCD_c176, CP-PACS_comp,SESAM_96} there is no
unambiguous sign for string breaking in simulations of QCD
with~$\nf=2$ flavours of Wilson fermions, if the potential is
determined by measuring Wilson loops. It has, however, been observed
for QCD with staggered quarks~\cite{Trottier_3D,MILC_sb_lat99} and in
QCD at finite temperature~\cite{Biele_sb}.

\begin{figure}[tb]
\vspace{0.8cm}
\hspace{-0.5cm}
\ewxy{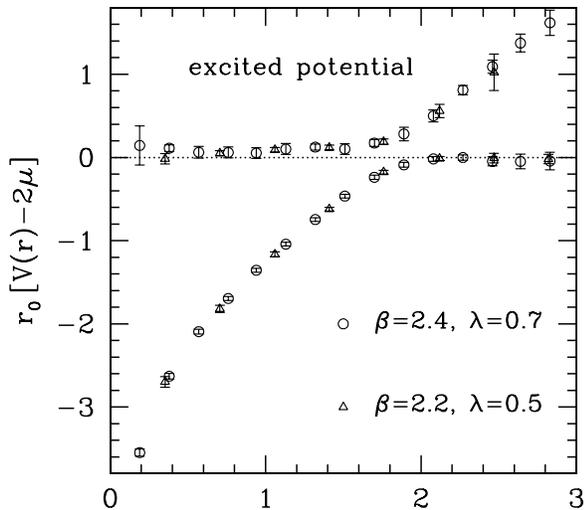}{90mm}
\vspace{-2.9cm}
\caption{The static quark potential and the first excitation computed
  in the SU(2) Higgs model using a variational
  technique~\protect\cite{KneSo}. The flattening of the ground state
  for large separations indicates the breakdown of linear confinement.
}
\label{fig_stringb}
\end{figure}

The reason for the failure to detect string breaking in QCD at zero
temperature has been attributed to the bad projection properties of
Wilson loops onto the state of the broken string. This has been
confirmed in studies in which the fermionic fields have been replaced
by scalar fields (which are computationally much less demanding whilst
preserving the underlying mechanism for string breaking to
occur)~\cite{string,KneSo,pws}. In these studies string breaking was
treated as a mixing phenomenon between the string and the
two-``meson'' state. This was achieved by supplementing the operator
basis with operators having an explicit projection onto the broken
string and by employing a variational approach to determine the energy
levels. Indeed this more sophisticated method has provided clear
evidence for string breaking (an example for the SU(2) Higgs model is
shown in Fig.~\ref{fig_stringb}), and there are preliminary
results~\cite{Utah_lat99,Pennanen_lat99} which indicate that the
approach is successful in QCD as well.

\section{Light quark masses}
\label{sec_quarkmass}

Quark masses are fundamental parameters of the Standard Model, and
their determination has been the subject of many recent activities. In
particular, the mass of the strange quark, $\ms$, has received a lot
of attention following the recent experimental results
on~$\epsilon^\prime/\epsilon$~\cite{eepr_exp}. Theoretical analyses of
this quantity rely on $\ms$ as one of the essential input parameters.

Ratios of the light quark masses are predicted by Chiral Perturbation
Theory with a precision of a few percent~\cite{Leutw96}. In order to
obtain individual quark masses it is thus sufficient to determine a
particular linear combination using lattice QCD.

\subsection{Non-perturbative renormalisation}

A convenient starting point to discuss lattice calculations of light
quark masses is the PCAC relation. For charged kaons it can be written
as
\be
  \fK\mK^2 = (\mbaru+\mbars) \langle0|\overline{u}\gamma_5{s}|
  {\rm K}\rangle.
\label{eq_PCAC}
\ee
In order to determine the sum of quark masses $(\mbaru+\mbars)$ using
the experimental result for $\fK\mK^2$, it suffices to compute the
matrix element $\langle0|\overline{u}\gamma_5{s}|{\rm K}\rangle$ in a
lattice simulation. The renormalisation of the pseudoscalar density
$\overline{u}\gamma_5{s}$ is, however, scale- and scheme-dependent. By
convention quark masses are quoted in the $\MSbar$ scheme of
dimensional regularisation at some reference scale~$\mu$. Therefore
the relation between the pseudoscalar densities in the $\MSbar$ scheme
and lattice regularisation must be computed, viz
\be
   \left(\overline{u}\gamma_5{s}\right)_{\MSbar} =
   \zp(g_0,a\mu)\,\left(\overline{u}\gamma_5{s}\right)_{\rm lat}.
\ee
Here $g_0$ is the bare coupling, and~$\mu$ is the subtraction point in
the $\MSbar$ scheme. The renormalisation factor~$\zp$ has been
computed in lattice perturbation theory for several discretisations.
However, the limitations of lattice perturbation theory are well
known, and in order to remove all doubts about the reliability of the
matching procedure it is evident that a non-perturbative determination
of the renormalisation factor is required.

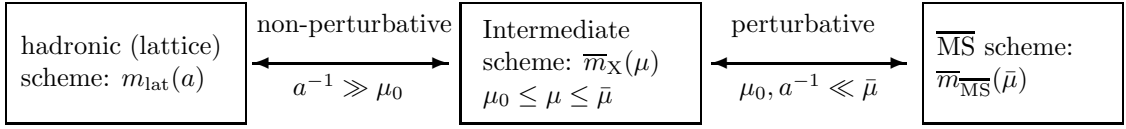
\begin{figure*}[t]
\begin{center}
\unitlength 1.06cm
\begin{picture}(14,2.0)
\put(0.0,0.0){
   \framebox(3.0,1.5){
     \parbox{28mm}{\sloppy
       hadronic (lattice)\\[0.3ex] scheme: $m_{\rm lat}(a)$
     }
  }
}
\put(5.7,0.0){
   \framebox(3.0,1.5){
     \parbox{26mm}{\sloppy
       Intermediate\\[0.3ex] scheme: $\mbar_{\rm X}(\mu)$
   \\[0.3ex] $\mu_0\leq\mu\leq\bar{\mu}$

     }
  }
}
\put(11.5,0.0){
   \framebox(2.5,1.5){
     \parbox{23mm}{\sloppy
       $\MSbar$ scheme:\\[0.3ex] $\mbar_{\MSbar}(\bar{\mu})$
     }
  }
}
\thicklines
\put(3.2,0.75){\vector(1,0){2.45}}
\put(5.6,0.75){\vector(-1,0){2.4}}
\put(3.25,1.15){\mbox{non-perturbative}}
\put(3.7,0.35){\mbox{${a}^{-1}\gg\mu_0$}}
\put(8.95,0.75){\vector(1,0){2.45}}
\put(11.35,0.75){\vector(-1,0){2.4}}
\put(9.2,1.15){\mbox{perturbative}}
\put(9.3,0.35){\mbox{$\mu_0,a^{-1}\ll\bar{\mu}$}}
\end{picture}
\end{center}
\caption{Schematic relation between quark masses in lattice
regularisation and the $\MSbar$ scheme through an intermediate
renormalisation scheme~$\rm X$.}
\label{fig_intscheme}
\end{figure*}

One of the main obstacles for a fully non-perturbative matching
procedure are the large scale differences between the low-energy
regime, which is the domain of the lattice regularisation scheme and
the high-energy, perturbative regime where the $\MSbar$ scheme is
defined. This technical difficulty can be overcome by introducing an
intermediate scheme~$\rm X$, as shown schematically in
Fig.~\ref{fig_intscheme}. The problem is thus split into two
parts. The first is the matching of the bare current quark mass
$m_{\lat}(a)$ to the running mass in the intermediate scheme,
$\overline{m}_{\rm X}(\mu)$. This amounts to
computing~$\zp(g_0,a\mu_0)$ between the lattice scheme and scheme~X
for a range of bare couplings~$g_0$ at a fixed scale~$\mu_0$. The
second part is the determination of the scale dependence of the
running mass~$\overline{m}_{\rm X}(\mu)$ from~$\mu_0$ up to very high
energies, where the perturbative relation
between~$\overline{m}_{\rm{X}}$ and~$\mbar_\MSbar$ can be expected to
be reliable. Through this two-step process the use of lattice
perturbation theory is completely avoided.

So far two proposals to define a suitable intermediate scheme have
been put forward. The first has been introduced in ref.~\cite{MPSTV94}
and imposes non-perturbative renormalisation conditions on Green
functions of local operators, computed between off-shell quark and
gluon states, with virtualities~$\mu$, in a fixed gauge. In the case
of the pseudoscalar density such a normalisation condition reads
\be
   \zp(g_0,a\mu)\,\zq^{-1}(g_0,a\mu)\,
   \Gamma_{\rm P}(ap)\Big|_{p^2=\mu^2} = 1.
\ee
Here $\Gamma_{\rm P}$ denotes the amputated Green function of
$\qbar\gamma_5{q}$, and $\zq$ is the quark wavefunction renormalisation
factor (which is easily computed in a lattice simulation). Typically
this relation is evaluated in the Landau gauge. This choice of
intermediate scheme is referred to as the Regularisation Independent
(RI) scheme, in which the r\^ole of the renormalisation scale is
played by the virtuality~$\mu$ of the quark states. The scale
dependence can then be probed by choosing different external momenta
for the quark fields used to compute $\Gamma_{\rm P}$. Provided that
the virtualities can be fixed such that
\be
   \Lambda_{\rm QCD}\ll\mu\ll{a^{-1}}
\ee
the perturbative matching between the RI and~$\MSbar$ schemes expected
to work well.

\begin{figure}[tb]
\vspace{-1.5cm}
\ewxy{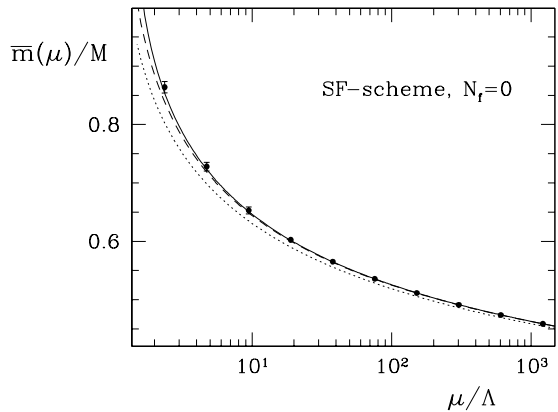}{98mm}
\vspace{-3.0cm}
\caption{Non-perturbative scale evolution of $\mbar_{\rm SF}/M$
computed in lattice simulations of the SF (solid circles). The dotted,
dashed and solid lines correspond to the scale evolution computed
using the 2/1-, 2/2 and 3/2-loop expressions for the RG
$\beta$-function and the anomalous dimension.}
\label{fig_mbarSF}
\end{figure}

Another intermediate scheme is defined using the Schr\"odinger
Functional (SF) of QCD~\cite{LNWW92,sint_SF,martin_LesHouches}. This
scheme is based on the formulation of QCD in a finite volume of
size~$L^3\cdot{T}$ with inhomogeneous (Dirichlet) boundary conditions
in the time direction. Non-perturbative renormalisation conditions can
then be imposed at scale~$\mu=1/L$ and zero quark mass. An attractive
feature of the SF scheme is that is allows to compute the scale
dependence non-perturbatively over several orders of magnitude, using
a recursive finite-size scaling technique. Thus, once the scale
dependence of $\mbar_{\rm SF}$ is known up to energies of around
100~GeV one can continue the scale evolution to infinite energy using
the perturbative renormalisation group functions and thereby extract
the renormalisation group invariant (RGI) quark mass~$M$. In
Fig.~\ref{fig_mbarSF} the scale dependence of $\mbar_{\rm SF}/M$
computed in quenched QCD~\cite{mbar1} is shown and compared to the
perturbative scale evolution.

The left-most point in Fig.~\ref{fig_mbarSF} corresponds to a scale
$\mu_0\sim275\,\mev$. Here one reads off~\cite{mbar1}
\be
   \frac{M}{\mbar_{\rm SF}}=1.157\pm0.015.
\label{eq_Mmbar}
\ee
The matching between lattice regularisation and $\MSbar$ scheme via
the SF is completed by computing the renormalisation
factor~$\zp(g_0,a\mu_0)$ for a particular fermion discretisation, at
fixed $\mu_0=275\,\mev$, for a range of bare couplings. In
ref.~\cite{mbar1} this has been performed for the $O(a)$ improved
Wilson action, for couplings which correspond to lattice spacings in
the range
\be
   a\sim0.1-0.045\,\fm.
\ee

\subsection{Light quark masses in quenched QCD}

Estimates for the light quark masses for several fermion lattice
actions and using non-perturbative renormalisation (implemented either
in the RI or SF schemes) have been reported
recently~\cite{JLQCD_quark_stag,mbar3,QCDSF_quark,Wing_lat99}. As an
illustration I shall describe in some detail the determination of the
strange quark mass along the lines of ref.~\cite{mbar3}, in which the
non-perturbative renormalisation factors computed in ref.~\cite{mbar1}
were used.

\begin{figure}[tb]
\vspace{-3.5cm}
\ewxy{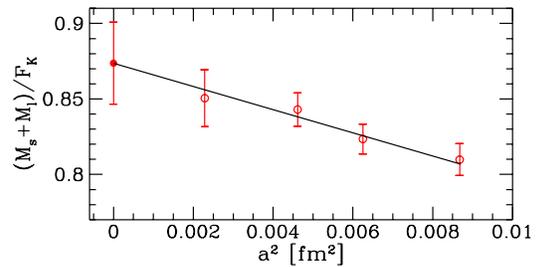}{98mm}
\vspace{-2.5cm}
\caption{Continuum extrapolation of $(M_{\rm s}+M_{\rm l})/{\fK}$. The
lattice scale is set by hadronic radius~$r_0$.}
\label{fig_MsMlcont}
\end{figure}

The average light quark mass $m_{\rm l}$ is defined as
\be
   m_{\rm l}=\textstyle\frac{1}{2}(m_{\rm u}+m_{\rm d}).
\ee
By combining the result for $M/\mbar_{\rm SF}$ from~\eq{eq_Mmbar} with
the factor $\zp(g_0,a\mu_0)$ and the PCAC relation~\eq{eq_PCAC} one
obtains an expression for the sum of RGI quark masses $(M_{\rm
s}+M_{\rm l})$ in units of the kaon decay constant
\be
   \frac{M_{\rm s}+M_{\rm l}}{\fK} = \frac{M}{\mbar_{\rm SF}}\,
   \frac{\mK^2}{\zp(g_0,a\mu_0)\,G_{\rm K}(a)} + O(a^2).
\label{eq_MfK}
\ee
Here, up to a small mass dependent factor which arises in the $O(a)$
improved Wilson theory, $G_{\rm K}$ is equal to the matrix element of
the pseudoscalar density evaluated at the kaon mass. At this point all
reference to the intermediate SF scheme is cancelled in the product
$M/(\mbar_{\rm SF}\zp)$. What remains to be specified is the
experimental value of $\mK^2$ expressed in units of some quantity
which sets the lattice scale (e.g. the hadronic
radius~$r_0$~\cite{rainer_r0,alpha_r0}). The results for $(M_{\rm
  s}+M_{\rm l})/{\fK}$ as obtained from~\eq{eq_MfK} can now be
extrapolated to the continuum limit. This is shown in
Fig.~\ref{fig_MsMlcont}, which also illustrates that the remaining
discretisation errors in the $O(a)$ improved theory are consistent
with a leading behaviour proportional to~$a^2$.

Using the experimental value $\fK=160\pm2\,\mev$ one obtains in the
continuum limit
\be
   M_{\rm s}+M_{\rm l} = 140\pm5\,\mev.
\label{eq_MsMlres}
\ee
This result can now be converted into $\mbars^\MSbar(\mu)$ at
$\mu=2\,\gev$. First one combines~\eq{eq_MsMlres} with the prediction
from Chiral Perturbation Theory~\cite{Leutw96}
\be
   M_{\rm s}/M_{\rm l} = 24.4\pm1.5.
\label{eq_MChPT}
\ee
By integrating the (4-loop) perturbative RG functions in the $\MSbar$
scheme one obtains
\be
   \mbar^{\MSbar}(\mu)/M = 0.7208\;\;\hbox{\rm at}\;\;\mu=2\,\gev.
\label{eq_mbarM_MSbar}
\ee
Finally, combining eqs.~(\ref{eq_MsMlres}),\,(\ref{eq_MChPT})
and~(\ref{eq_mbarM_MSbar}) yields the final result in the quenched
approximation ~\cite{mbar3}
\be
   \mbars^\MSbar(2\,\gev) = 97\pm4\,\mev.
\label{eq_ms_alpha}
\ee
The quoted uncertainty of $\pm4\,\mev$ contains all errors, except
those due to quenching. As mentioned in the introduction, the
conversion into physical units is ambiguous in the quenched
approximation. For $\mbars^\MSbar(2\,\gev)$ the resulting uncertainty
was estimated to amount to $\sim10\,\%$.

\begin{table*}
\begin{center}
\caption{Estimates for the quark masses $m_{\rm s}$ and $m_{\rm l}$ in
  the $\MSbar$ scheme at $\mu=2\,\gev$ in quenched QCD. Also shown is
  the choice of intermediate renormalisation scheme~X and the lattice
  Dirac operator. Crosses indicate that non-perturbative
  renormalisation and/or a continuum extrapolation has not been
  implemented.
}
\label{tab_quarkmass}
\begin{tabular}{lllclc}
\hline\hline\\[-0.5ex]
Collab. & $m_{\rm l}$ & $m_{\rm s}$ & X & $\Sf$ & $a\to0$ \\[0.5ex]
\hline\\[-0.5ex]
BGLM \cite{BGLM99}      & 4.5(5)  & 111(9)      & RI    & $D_{SW}$
                        & {$\times$} \\
QCDSF \cite{QCDSF_quark}& 4.4(2)  & 105(4)      & SF    & $D_{SW}$
                        & {$\surd$} \\
RBC \cite{RBC_DWF}      &         & 130(11)(18) & RI    & $D^{\rm DWF}$
                        & {$\times$} \\
ALPHA/UKQCD \cite{mbar3}&         & ~97(4)      & SF    & $D_{SW}$
                        & {$\surd$} \\
CP-PACS \cite{CP-PACS_quen}& 4.55(18) & 115(2)  & $\times$ & $D_{W}$
                        & {$\surd$} \\
BSW \cite{BSW_quark}    &         & ~96(26)     & $\times$ & $D^{\rm DWF}$
                        & {$\surd$} \\
JLQCD \cite{JLQCD_quark_stag}& 4.23(29) & 106(7)& RI    & $D_{\rm stag}$
                        & {$\surd$} \\
Becirevic               &         &             &       &       & \\
et al. \cite{Becir98}   & \rb{4.5(4)}  & \rb{111(12)}   & \rb{RI} &
                        \rb{$D_{SW}$}  & \rb{{$\times$}} \\[0.5ex]
\hline\hline
\end{tabular}
\end{center}
\end{table*}

A compilation of recent results for $m_{\rm s}$ and $m_{\rm l}$ in
quenched QCD is shown in Table~\ref{tab_quarkmass}. When comparing the
results one has to bear in mind that systematic errors -- where shown
-- have not been estimated in a uniform manner, or have sometimes been
combined in quadrature with the statistical errors. Also, the
conversion into physical units has been performed using different
quantities. Nevertheless, the picture that emerges is quite
encouraging. Estimates for the strange quark mass in quenched QCD
cluster around 100\,\mev\ and around 4.5\,\mev\ for the average up and
down quark mass. Different discretisations such as Wilson, staggered
and Domain Wall fermions as well as different implementations of
non-perturbative quark mass renormalisation yield broadly consistent
results.

\subsection{Sea quark effects in $m_{\rm s}$ and $m_{\rm l}$}

An important issue, especially for phenomenological applications of
lattice calculations of the light quark masses is the influence of
dynamical quarks. Estimates for $m_{\rm l}$ and $\ms$ computed in
partially quenched QCD have been reported
in~\cite{SESAM_quark97,SESAM_nf2_98,kaneko_lat99,cb_quark_lat99}.
Since non-perturbative renormalisation has so far not been applied for
$\nf=2$, all of these calculations rely on perturbation theory to
match the quark masses in the lattice and $\MSbar$ schemes.

\begin{figure}[tb]
\vspace{-1.5cm}
\hspace{-0.8cm}
\ewxy{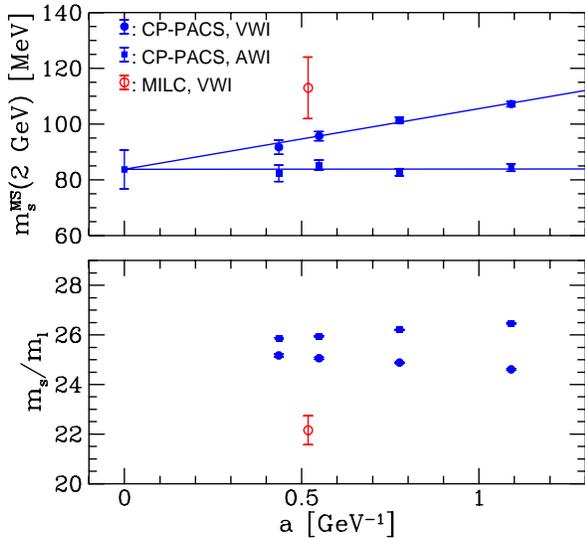}{115mm}
\vspace{-3.2cm}
\caption{The strange quark mass and the ratio $\ms/m_{\rm l}$
extracted using either the vector or axial vector Ward identities
plotted versus the lattice spacing.}
\label{fig_mstr_nf2}
\end{figure}

The most comprehensive study so far has been presented by
CP-PACS~\cite{kaneko_lat99} in which not only the $\msea$-dependence
has been investigated but also the extrapolation to the continuum
limit.  The scaling behaviour of $\mbars^{\MSbar}(2\,\gev)$ extracted
either from the vector or axial vector Ward identities (labelled VWI
or AWI respectively) is shown in the upper part of
Fig.~\ref{fig_mstr_nf2}.  In spite of the large differences observed
for the two methods at non-zero lattice spacing, the results are
consistent with a common continuum limit, where
$\mbars^{\MSbar}(2\,\gev)=84\pm7\,\mev$. This is significantly lower
than the quenched results discussed earlier.  However, many systematic
effects are not as well controlled as in the quenched case, so that
further studies, employing non-perturbative renormalisation and
investigating the $\nf$-dependence are required before a substantial
decrease of quark masses relative to the quenched results can be
confirmed. It is interesting, though, that the ratio $\ms/m_{\rm l}$,
in which many systematic effects are expected to cancel, extrapolates
to $\ms/m_{\rm l}=26\pm3$ (c.f. lower part of
Fig.~\ref{fig_mstr_nf2}). This value is in excellent agreement with
the prediction from Chiral Perturbation Theory,~\eq{eq_MChPT}. The
preliminary results by the MILC Collaboration~\cite{cb_quark_lat99},
which are also shown in the figure, have been computed at a fixed
value of~$\msea$ corresponding to $m_\pi/m_\rho=0.56$. This may
explain why they are different from the CP-PACS results, which in turn
have been extrapolated in~$\msea$ to the physical value.

For phenomenological applications one may be tempted to convert the
results presented here into a global estimate. Before more thorough
studies of dynamical quark effects become available, I consider the
quenched results (e.g. \eq{eq_ms_alpha}) the most reliable estimate.
By accounting for the systematic errors due to using the quenched
approximation one can then quote a global result for $\ms$ as
\be
   \mbars^{\MSbar}(2\,\gev) = 100\pm5\,(\rm stat)\,^{+10}_{-20}\,(\rm
   syst)\,\mev.
\ee
The systematic error in the above estimate incorporates the
afore-mentioned scale ambiguity of 10\,\mev, as well as the observed
decrease in the central value for $\nf=2$ dynamical flavours.

\section{Glueballs and heavy hybrids}
\label{sec_glueb}

Historically glueball masses were among the first quantities to be
computed in lattice gauge theories. Over the years the calculations
have become more refined, for instance, by constructing efficient
glueball operators, by studying the approach to the continuum limit,
and by including higher spin states.

The main difficulty in glueball calculations is the relatively high
level of statistical noise in the correlation functions from which the
masses are extracted. Here I shall focus on recent calculations in
which this problem has been alleviated by using anisotropic lattice
actions. Hence, from now on I shall distinguish between spatial and
temporal lattice spacings, $a_s$ and~$a_t$, respectively. If $C_{\rm
G}(t)$ denotes the correlation function of a glueball operator $G(x)$,
then its asymptotic behaviour for large separations~$t$ is given by
\be
  C_{\rm G}(t)=\sum_{\vec{x}}\,\langle G(\vec{x},t)G^\dagger(0)\rangle
  \sim \rme^{-(a_tM_{\rm G})(t/a_t)}.
\ee
That is, the exponential decay of~$C_{\rm G}(t)$ is governed by the
glueball mass in units of the temporal lattice spacing. Therefore, the
larger $a_tM_{\rm G}$, the quicker the decay of $C_{\rm G}(t)$, so
that its asymptotic behaviour may be difficult to isolate before the
statistical noise becomes too large. By introducing an anisotropic
lattice action with $a_t\ll a_s$, one can simultaneously achieve slow
exponential fall-off of $C_{\rm G}(t)$ whilst preserving large spatial
volumes in physical units. The spatial volume in {\it lattice units\/},
however, can be kept small if $a_s$ is large in physical units, so
that the calculations are more manageable. Typical values of~$a_s$ are
\be
   a_s \approx 0.2-0.4\,\fm,
\ee
while the ``aspect ratio''~$\xi\equiv a_s/a_t$ is usually taken in
the range
\be
   \xi=3-5.
\ee
The idea of using anisotropic lattices in glueball calculations is not
new: it was used already~16 years ago~\cite{Mike_aniso}, before the
advent of ``smearing'' techniques to construct glueball operators for
which the asymptotic behaviour in the correlation function sets in
very quickly.

The idea of anisotropic lattice actions has been revived by the desire
to compute the higher glueball states more accurately and to use very
coarse spatial lattices so that these calculations can be performed on
smaller computers. However, by pushing to very large values of~$a_s$
one may suffer from uncontrollably large discretisation effects. Here,
the Symanzik improvement programme can be used to eliminate the
leading lattice artefacts, so that the extrapolation of results
obtained on coarse, anisotropic lattices is sufficiently controlled.

In the rest of this section I shall mainly discuss recent results for
the spectra of glueballs and heavy hybrids obtained using anisotropic
actions. For more comprehensive reviews I refer the reader to
refs.~\cite{Teper_98,Toussaint_lat99}.

\subsection{Glueball spectrum in quenched QCD}

\begin{figure}[tb]
\vspace{0.3cm}
\ewxy{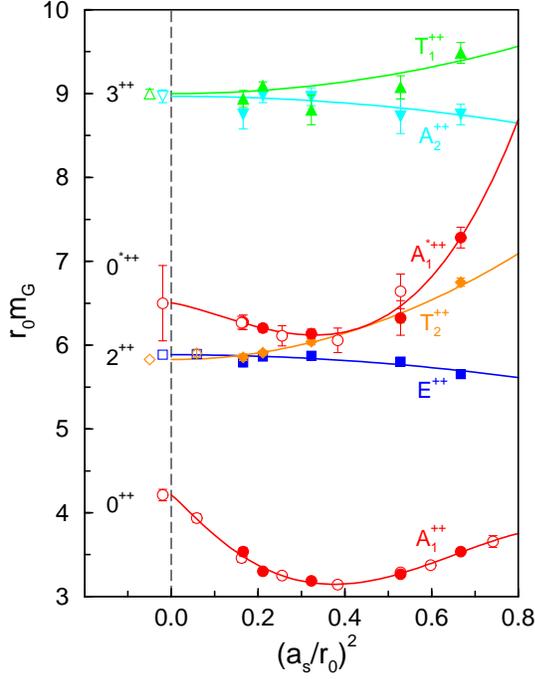}{90mm}
\vspace{-0.7cm}
\caption{Continuum extrapolations of glueballs computed for different
representations of the octahedral group on anisotropic
lattices~\protect\cite{MorPea} for ${PC}=++$.}
\label{fig_r0mG_ext}
\end{figure}

In two recent papers Morningstar and Peardon~\cite{MorPea} have
presented results for the quenched glueball spectrum below $4\,\gev$
computed using an anisotropic, $O(a_s^2)$ improved gluon lattice
action with aspect ratios $\xi=3$ and~5. Thus, improvement has been
employed to reduce artefacts associated with the spatial lattice
spacing only. Non-perturbative determinations of the relevant
improvement coefficients are not available for the case at hand, so
that one relies on their estimates in mean-field improved perturbation
theory. This procedure does not completely remove lattice artefacts of
order~$a_s^2$. One thus expects leading cutoff effects of order
\be
   g^2a_s^2,\;a_s^4,\;a_t^2.
\ee
Since the lattice breaks rotational symmetry, glueball operators are,
as usual, constructed from representations of the octahedral group:
$A_1, A_2, E, T_1, T_2$. Figure~\ref{fig_r0mG_ext} shows the scaling
behaviour of the glueball masses in the $PC=++$ channel, extracted
from the various representations. The plot nicely demonstrates the
restoration of rotational symmetry in the limit $a_s\to0$. For
instance, both the~$E$ and~$T_2$ representations describe the spin-2
glueball in the continuum limit, but differ significantly at large,
non-zero lattice spacings.

\begin{figure}[tb]
\vspace{-0.5cm}
\ewxy{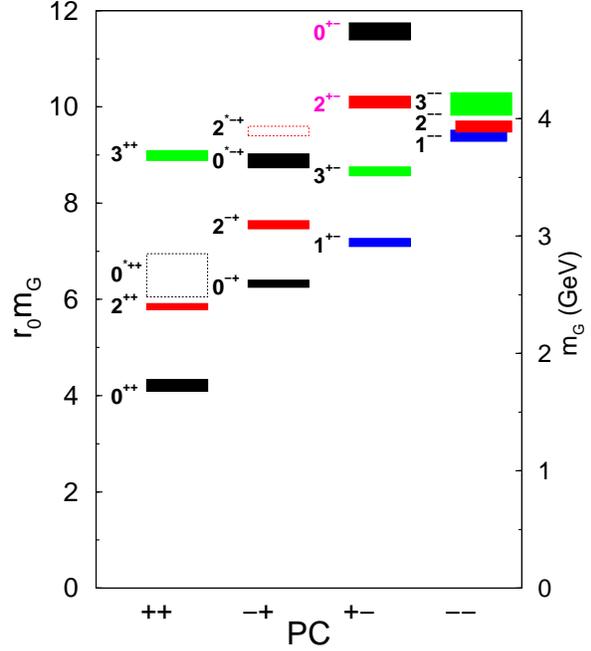}{100mm}
\vspace{-0.9cm}
\caption{Results for the quenched glueball spectrum in the continuum
limit from ref.~\protect\cite{MorPea}.}
\label{fig_mG_cont}
\end{figure}

The observed curvature in the continuum extrapolation of the scalar
($0^{++}$ and ${0^\ast}^{++}$) glueballs suggests that both $O(a_s^2)$
and $O(a_s^4)$ lattice artefacts are sizeable for these states. Hence,
if lattice spacings as large as $a_s\approx0.4\,\fm$ are included one
has to use a more complicated model function for the continuum
extrapolation, which ultimately leads to a loss of statistical
precision for these states.

The final, continuum results for glueball states from
ref.~\cite{MorPea} for a number of different $J^{PC}$ assignments are
shown in Fig.~\ref{fig_mG_cont}, in units of the hadronic radius~$r_0$
and in physical units on the left and right margins, respectively.

The masses of the two lowest-lying glueballs obtained on anisotropic
lattices can now be compared to results using more conventional
techniques. A meaningful comparison can be made by expressing results
from different simulations in units of a common scale, for which I
have again chosen the hadronic radius~$r_0$. Thus, the results from
refs.~\cite{UKQCD_glueb93,Teper_98,GF11_glueb99} have been expressed
or converted in units of~$r_0$, and whenever necessary the continuum
extrapolation has been re-done. The resulting masses for the lowest
scalar and tensor glueballs are listed in Table~\ref{tab_mG_comp}
together with the value of the aspect ratio~$\xi$ and the year in
which the calculation was carried out. It is also indicated whether
a continuum extrapolation has been performed.

\begin{table*}
\begin{center}
\caption{Comparison of the two lowest glueball masses in units
of~$r_0$.} 
\label{tab_mG_comp}
\begin{tabular}{lllccc}
\hline\hline\\[-0.5ex]
Collab.  &  $r_0\,m_{0^{++}}$ & $r_0\,m_{2^{++}}$ & $\xi$ & $a\to0$ &
year \\[0.5ex]
\hline\\[-0.5ex]
M+P\,\cite{MorPea}          & 4.21(11)(4) & 5.85(2)(6) & 3,\,5
 & $\surd$ & 1999 \\
GF11\,\cite{GF11_glueb99}   & 4.33(10)    & 6.04(18) & 1
 & $\surd$ & 1999\\ 
Teper\,\cite{Teper_98}      & 4.35(11)    & 6.18(21) & 1
 & $\surd$ & 1998\\ 
UKQCD\,\cite{UKQCD_glueb93} & 4.05(16)    & 5.84(18) & 1
 & $\times$ & 1993\\ [0.5ex]
\hline
\hline
\end{tabular}
\end{center}
\end{table*}

The table shows that the results obtained in different simulations are
in broad agreement. If those estimates are to be used for
phenomenological purposes, it should be kept in mind that they are
valid in the quenched approximation. This implies that the conversion
into physical units (e.g. by using $r_0=0.5\,\fm$) is ambiguous. Also,
the influence of dynamical quark effects, and, perhaps most
importantly, the issue of glueball-meson mixing has not been addressed
for the results presented in Table~\ref{tab_mG_comp}. The latter has
been studied, e.g. in ref.~\cite{Teper_NATO97,Amsler_Close,GF11_mix},
and very recently in ref.~\cite{GF11_mix99}, which also contains a
detailed calculation of the spectrum of light quarkonia in quenched
QCD.  Overall, it is found that physical glueball masses are quite
sensitive to the details of the proposed mixing pattern.

Dynamical quark effects in the glueball spectrum have been studied
in~\cite{SESAM_glueb_nf2} and~\cite{cmi_disc}. Whereas
ref.~\cite{SESAM_glueb_nf2} reports no significant deviation of
glueball masses computed for $\nf=2$ (within large errors), results by
UKQCD~\cite{cmi_disc} indicate much lower estimates in unquenched
simulations for both the scalar glueball and quarkonium
state. However, a number of effects, e.g. the question of lattice
artefacts have to be addressed in much more detail before these
results can be confirmed.

\subsection{Heavy quarkonia and hybrids}

In addition to glueballs the spectrum of quarkonia and hybrids can also
be calculated in lattice QCD. Such calculations may provide hints
especially for the heavy quark sector, where very little experimental
data exists for hybrid states. Thus, on the lattice one would like to
compute the masses of states which are obtained from quark bilinears
by inserting one or more gluons, such as $\overline{b}gb$ and
$\overline{c}gc$.

Ideally a fully relativistic treatment of the heavy quarks would be
desirable. However, for the currently accessible range of lattice
spacings, for which $a\;\gtaeq\;0.05\,\fm$ (corresponding to
$a^{-1}\;\lesssim\;4\,\gev$) one expects large lattice artefacts in
the charm sector, while the $b$ quark with a mass above $4\,\gev$
cannot at all be simulated directly. If anisotropic lattices with
coarse spatial lattice spacings of~$a_s=0.2-0.4\,\fm$ are employed,
the situation is even worse.

One way to address the problem of large cutoff effects in the heavy
quark sector is to resort to an effective, non-relativistic
treatment~\cite{NRQCD}, by introducing a cutoff~$\Lambda$ such that
\be
   \Lambda\sim a^{-1} < m_Q,
\label{eq_relcut}
\ee
where $m_Q$ is the mass of the heavy quark. In other words,
relativistic states above~$\Lambda$ are excluded. Based on this
approximation one can write down a discretised, effective,
non-relativistic QCD action (NRQCD). On a formal level this
approximation of QCD can be viewed as an expansion in the strong
coupling constant and the 4-velocity~$v$ of the heavy quark. In this
formulation the lattice spacing acts both as the UV regulator and the
non-relativistic cutoff. This implies in turn that the continuum limit
$a\to0$ cannot be taken, since otherwise the non-relativistic
approximation breaks down. Therefore, in order for cutoff effects to
be under control one relies on a ``window'' in~$a$ where NRQCD works
well and lattice artefacts are small at the same time.

Two groups have recently reported results for heavy quarkonia and
hybrids using NRQCD on anisotropic
lattices~\cite{CP-PACS_hybrids,KJM_hybrids}. Both have used the
$O(a_s^2)$ mean-field improved gluon action for $\xi=3,\,5$ and an
NRQCD action expanded in the 4-velocity to order $m_Qv^2$. At this
order, spin interactions are neglected. Therefore one expects spin
degeneracies for the $S$- and $P$-wave quarkonium states, as well as
for each of the hybrid states $H_1$, $H_2$ and $H_3$ with quantum
numbers $J^{PC}=1^{--}$ ($H_1$), $1^{++}$ $(H_2)$ and $0^{++}$
($H_3$), respectively. 

\begin{figure}[tb]
\vspace{-1.3cm}
\hspace{-1.0cm}
\ewxy{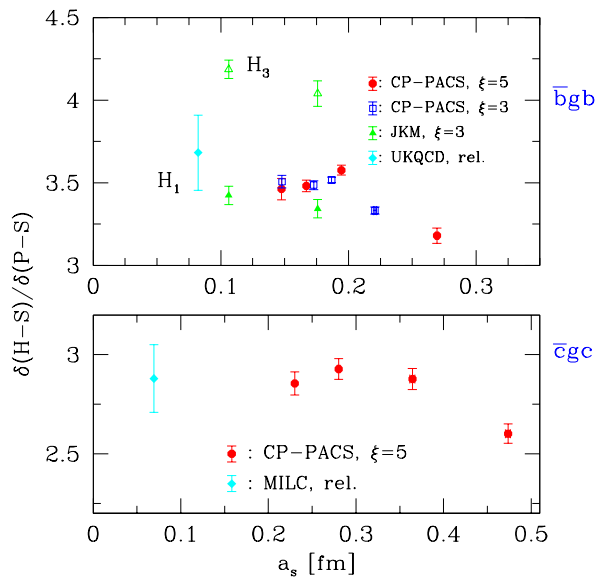}{115mm}
\hspace{0.0cm}
\vspace{-3.2cm}
\caption{The hybrid splittings in the charm and bottom sectors
  from NRQCD on anisotropic lattices. Diamonds denote the results
  using relativistic heavy quarks.}
\label{fig_hybrids}
\end{figure}

A convenient way to present the results is to quote the
hybrid-$S$-wave splittings $\delta(H_i-S), i=1,2,3$. In
Fig.~\ref{fig_hybrids} those splittings, normalised to the $1S-1P$
quarkonium splitting, are plotted versus the spatial lattice
spacing~$a_s$. Focussing on the upper part of the figure which shows
the spectrum in the bottomonium sector, one sees that anisotropy
effects appear to be under control: the results by
CP-PACS~\cite{CP-PACS_hybrids} obtained for $\xi=3,5$ are consistent
within errors. Furthermore, the data from both groups support the
existence of a window for $a_s\approx0.1-0.2\,\fm$, where
discretisation effects are small: in that range the variation of the
results with $a_s$ is roughly as large as the statistical precision.
It is obvious, though, that this is no longer the case for $a_s>
0.2\,\fm$. Similar observations apply in the charmonium sector shown
in the lower part of Fig.~\ref{fig_hybrids}, where the scaling window
appears to be somewhat larger.

The results for the lowest $\overline{b}gb$ hybrid splitting obtained
on anisotropic lattices are~\cite{CP-PACS_hybrids,KJM_hybrids}
\be
    \delta(H_1-S) = \left\{\begin{array}{ll}
        1.542(8)\,\gev   & \hbox{CP-PACS} \\
        1.49(2)(5)\,\gev & \hbox{KJM}
        \end{array} \right.
\ee
For both charmonium and bottomonium hybrids the results can be
compared to those in which the heavy quarks were treated
relativistically, involving extrapolations in the heavy quark
mass~\cite{UKQCD_hyb_rel,MILC_hyb_rel}. These data points are included
as diamonds in Fig.~\ref{fig_hybrids} and are consistent with those
using anisotropic actions. This nicely illustrates how complementary
formulations of heavy quarks can be used to control different
systematic effects.

\section{Other topics}
\label{sec_other}

The topics which are presented in this section could not be reviewed
extensively, but nevertheless I shall briefly summarise the current
status and refer the interested reader to the recent literature.

\subsection{Kaon weak matrix elements}

Lattice calculations for the $B$-parameter $B_K$, which is relevant for
$K^0-\overline{K}^0$ mixing, as well as matrix elements for
$K\to\pi\pi$ and $\epsilon^\prime/\epsilon$ have recently been
reviewed by Kuramashi~\cite{Kura_lat99} and
Martinelli~\cite{Marti_kaon99}. 

There are now many calculations of $B_K$ using different
discretisations of the Dirac operator. For staggered fermions the most
recent result quoted in the naive dimensional reduction (NDR) scheme
is
\be
   B_K^{\rm NDR}(2\,\gev) = 0.628\pm0.048.
\label{eq_BK_KS}
\ee
Further improvements of this result could be achieved through the
implementation of non-perturbative renormalisation. Simulations using
Wilson fermions~\cite{BK_Wilson} are consistent with the result in
\eq{eq_BK_KS}. Estimates for $B_K$ using Domain Wall fermions have
also been presented~\cite{BluSo_BK,BluSo_lat99}. 

Compared to $K^0-\overline{K}^0$ mixing the lattice predictions for
matrix elements relevant for $K\to\pi\pi$ and
$\epsilon^\prime/\epsilon$ are a lot less accurate, and a number of
systematic effects have to be better controlled. Details can be found
in~\cite{Kura_lat99,Marti_kaon99}. A surprising, {\it negative\/}
result for $\rm Re(\epsilon^\prime/\epsilon)$ has been reported
recently in~\cite{BRC_eps}. However, given that it has been obtained
using the relatively new technology of Domain Wall fermions, future
studies are required to clarify this issue.

\subsection{$B$ decay matrix elements}

Decays of heavy-light mesons have long been  studied in lattice gauge
theories. The current status has been reviewed by
Hashimoto~\cite{Hashi_lat99}. Semi-leptonic heavy-to-light and
heavy-to-heavy decays have been analysed in the quenched approximation
for several formulations of heavy quarks. Lattice estimates for
heavy-light decay constants such as $f_B$ have stabilised in the
quenched approximation, and their current values can be summarised
as~\cite{Hashi_lat99}
\be
  f_B=170\pm20\,\mev,\;f_{B_s}/f_B=1.15\pm0.04.
\ee
Present activities center on the quantification of sea quark effects.
Preliminary estimates suggest that heavy-light decay constants
increase by up to 20\,\% for $\nf=2$ flavours of dynamical quarks at
non-zero values of the lattice spacing~\cite{Hugh_lat99}. However, a
detailed scaling analysis is still required in order to separate sea
quark effects from lattice artefacts, so that the large increase in
$f_B$ for $\nf=2$ does, in my view, not constitute a solid result at
present.

\subsection{Structure functions}

This topic has recently received a lot of attention and has been
reviewed by Petronzio~\cite{Petro_lat99}. Current activities include
the implementation of non-perturbative renormalisation of parton
density operators, using either the RI or SF schemes described in
section~\ref{sec_quarkmass}. The SF scheme has been employed to
determine non-perturbatively the scale dependence of the twist-two,
non-singlet parton density operator~\cite{GuJaPet_99}. Other projects
in this area include calculations of higher twist contributions to the
pion structure function~\cite{QCDSF_tw4}.

\section{Summary}
\label{sec_summary}

The most significant theoretical development in Lattice Gauge Theory has
surely been the progress made in formulating chiral symmetry at
non-zero lattice spacing. It is now clear that chiral gauge theories
can be put on the lattice in a consistent way, without breaking the
gauge symmetry. This shows that a regularisation of the Standard Model
exists beyond perturbation theory.

Lattice simulations of QCD are becoming ever more refined, thanks to a
number of technical developments, such as the implementation of the
Symanzik improvement programme, non-perturbative renormalisation, the
use of anisotropic lattices and bigger, more efficient simulations
with dynamical quarks. 

The quenched approximation, which is still widely used, works
surprisingly well. This is good news for many computationally more
demanding applications, for which the quenched approximation will be
useful for some time in the future. There are now many attempts to
quantify the effects of dynamical quarks, and in some cases they have
been found to be significant. This has only been possible after all
other systematic effects, including lattice artefacts, could be
controlled at the level of a few percent.

Although lattice QCD may not  yet be the ultimate phenomenological
tool, it is clear that enormous progress towards this goal has been,
and will be made.

\vspace{0.7cm}

\par\noindent
{\bf Acknowledgements} \\[0.2cm]
%
%
I am grateful to Arifa Ali Khan, Oliver B\"ar, Claude Bernard, Ruedi
Burkhalter, Chris Dawson, Takashi Kaneko, Thomas Manke, Colin
Morningstar, Petrus Pennanen, Gerrit Schierholz, Hugh Shanahan,
Amarjit Soni and Matt Wingate for making their results available to
me. I thank Sinya Aoki, Ruedi Burkhalter, Michael Haas, Shoji
Hashimoto, Kazuyuki Kanaya and Frithjof Karsch for useful discussions,
and Karl Jansen for a critical reading of the manuscript. I am
grateful to Akira Ukawa for the kind hospitality at the Center for
Computational Physics, University of Tsukuba, where most of this talk
was prepared. The support by PPARC through the award of an Advanced
Fellowship is gratefully acknowledged.

\end{document}